\documentclass[useAMS,usenatbib]{mn2e}

\usepackage{graphicx}

\begin{document}


\newcommand{\lya}{Ly$\alpha$}
\newcommand{\heii}{He\textsc{ii}}
\newcommand{\hii}{H \textsc{ii}}
\newcommand{\hei}{He\textsc{i}}
\newcommand{\ion}[2]{[\textsc{#1}]#2}
\newcommand{\ha}{\ensuremath{\mathrm{H}\alpha}}
\newcommand{\hb}{\ensuremath{\mathrm{H}\beta}}
\newcommand{\nii}{{\ion{N ii}{6564}}}
\newcommand{\sii}{{\ion{S ii}{6716,6731}}}
\newcommand{\siii}{{\ion{S iii}{9069,9532}}}
\newcommand{\oiii}[1][5007]{{\ion{O iii}{#1}}}
\newcommand{\oii}{{\ion{O ii}{3727}}}
\newcommand{\ewha}{\ensuremath{\mathrm{EW(\ha)}}}
\newcommand{\dewha}{\ensuremath{\Delta \mathrm{EW}(\ha)}}
\newcommand{\ssfr}{SFR/M$_*$}
\newcommand{\HII}{H\textsc{ii}}
\newcommand{\apjs}{ApJS}
\newcommand{\apj}{ApJ}
\newcommand{\aj}{AJ}
\newcommand{\araa}{ARA\&A}
\newcommand{\apjl}{ApJL}
\newcommand{\nat}{Nature}
\newcommand{\pasp}{PASP}
\newcommand{\aap}{A\&A}
\newcommand{\mnras}{MNRAS}

\def\ltsima{$\; \buildrel < \over \sim \;$}
\def\simlt{\lower.5ex\hbox{\ltsima}}
\def\gtsima{$\; \buildrel > \over \sim \;$}
\def\simgt{\lower.5ex\hbox{\gtsima}}

\newcommand{\Om}{\Omega_{\rm m}}
\newcommand{\Omo}{\Omega_{\rm m,0}}
\newcommand{\Ob}{\Omega_{\rm b}}
\newcommand{\Obo}{\Omega_{\rm b,0}}

\newcommand{\beq}{\[}
\newcommand{\eeq}{\]}
\newcommand{\bneq}{\begin{equation}}
\newcommand{\eneq}{\end{equation}}


\title[D/H in the most metal-poor DLA]{Deuterium Abundance in the
Most Metal-Poor Damped Lyman alpha System: Converging on
$\Omega_{\rm b,0} h^2$
}

\author[M. Pettini et al.]
       {Max Pettini$^1$, Berkeley J. Zych$^1$, Michael T. Murphy$^{2, 1}$,
        Antony Lewis$^1$ and
        \newauthor Charles C. Steidel$^3$\\
        $^1$Institute of Astronomy, University of Cambridge,
         Madingley Road, Cambridge CB3 0HA, UK\\
         $^2$Centre for Astrophysics and Supercomputing, Swinburne University
         of Technology, Mail H39, PO Box 218, Victoria 3122, Australia\\
	 $^3$California Institute of Technology, Mail Stop 105-24, Pasadena, CA 91125, USA
      }
\date{Accepted ---; Received ---; in original form ---}
\pagerange{\pageref{firstpage}--\pageref{lastpage}}
\pubyear{2008}

\maketitle

\label{firstpage}

\begin{abstract}

The most metal-poor DLA known to date, at $z_{\rm abs} = 2.61843$ in
the spectrum of the QSO Q0913+072, with an oxygen abundance only
$\sim 1/250$ of the solar value, shows six well resolved D\,{\sc i}
Lyman series transitions in high quality echelle spectra recently obtained
with the ESO VLT.
We deduce a value of the deuterium abundance
$\log {\rm (D/H)} = -4.56 \pm 0.04$ which is in
good agreement with four out of the six most reliable
previous determinations of this ratio in QSO absorbers.
We find plausible reasons why in the other two cases the
$1 \sigma$ errors may have been underestimated by
about a factor of two.
The addition of this latest data point does not
change significantly the mean value of the primordial
abundance of deuterium, suggesting that we are now
converging to a reliable measure of this quantity.
We conclude that
$\langle \log {\rm (D/H)_p} \rangle = -4.55 \pm 0.03$
and $\Omega_{\rm b,0}h^2{\rm  (BBN) } = 0.0213 \pm 0.0010$
(68\% confidence limits).
Including the latter as a prior in the analysis of the
five year data of \textit{WMAP} leads to a revised
best-fitting value of the power-law index of
primordial fluctuations
$n_{\rm s} = 0.956 \pm 0.013$ ($1 \sigma$)
and $n_{\rm s} < 0.990$ with 99\%  confidence.
Considering together the constraints provided by
\textit{WMAP~5}, (D/H)$_{\rm p}$, baryon oscillations
in the galaxy distribution, and distances to Type~Ia supernovae,
we arrive at the current best estimates
$\Omega_{\rm b,0}h^2 = 0.0224 \pm  0.0005$ and
$n_{\rm s} = 0.959 \pm 0.013$.
\end{abstract}


\section{Introduction}
\label{sec:introduction}

The exquisite precision with which the temperature anisotropies
of the cosmic microwave background (CMB) have been mapped
on the sky has allowed the determination of
cosmological parameters to better than 10\% (Dunkley et al. 2008).
It is still important, however, to measure these
parameters with alternative methods, partly
as a consistency check on the standard cosmological model,
but also because the
CMB fluctuations generally constrain \emph{combinations}
of more than one parameter (e.g. Bridle et al. 2003).
The contribution of baryons to the present-day critical density,
$\Obo h^2$
where as usual $h = H_0/100$\,km~s$^{-1}$~Mpc$^{-1}$,
is one such example: the value
of $\Obo h^2$ which best fits the
power spectrum of CMB fluctuations
is tied to other parameters, such as
the spectral index of primordial density perturbations,
$n_{\rm s}$, and the optical depth to reionisation, $\tau$
(see, for example, Pettini 2006).

The quantity $\Obo h^2$ can also be deduced from
the primordial abundances of the light elements whose
nucleosynthesis is the result of
physical processes that are entirely different
from the acoustic oscillations of the
photon-baryon fluid imprinted on the CMB, and that took place
at much earlier times---only a few hundred seconds, rather than
a few hundred thousand years, after the Big Bang.
Among the light elements produced by primordial
nucleosynthesis, deuterium is the one whose
abundance by number relative to hydrogen,
(D/H)$_{\rm p}$, depends
most sensitively on $\Obo h^2$, and is thus
the `baryometer' of choice
(Steigman 2007; Molaro 2008).

Reliable measures of (D/H)$_{\rm p}$ are difficult to obtain, however.
The astrophysical environments which seem most appropriate are the
hydrogen-rich clouds absorbing the light of background QSOs
at high redshifts, but rare combinations of: (a) neutral hydrogen
column density in the range
$17 \simlt \log [N$(H\,{\sc i})/cm$^{-2}] \simlt 21$;
(b) low metallicity [M/H] corresponding
to negligible astration of D; and (c) most importantly, low internal
velocity dispersion of the absorbing atoms allowing the isotope shift
of only 81.6\,km~s$^{-1}$ to be adequately resolved, are required
for this observational test to succeed.
Thus, while the potential of this method was appreciated
more than thirty years ago (Adams 1976) and first realised
with the advent of 8-10\,m class telescopes in the 1990s
(Tytler et al. 1995), the number of trustworthy measurements
of (D/H)$_{\rm p}$ is still only five or six (O'Meara et al. 2006;
Steigman 2007).

There are a strong incentives to increase these meagre statistics.
On the one hand,
the mean $\langle {\rm (D/H)_p} \rangle = (2.82\pm 0.26) \times 10^{-5}$
implies $\Obo h^2 {\rm (BBN)} = 0.0213 \pm 0.0013$ (O'Meara et al. 2006)
which agrees, within the errors,
with the value
$\Obo h^2 {\rm (CMB)} = 0.02273 \pm 0.00062$
obtained by Dunkley et al. (2008) from the analysis of five
years of observations with the \textit{Wilkinson
Microwave Anisotropy Probe (WMAP 5)}.
On the other hand, the standard deviation among the six
measures of (D/H)$_{\rm p}$ in high redshift
QSO absorbers exceeds the dispersion expected
from the individual errors (leading to a high value
of $\chi^2$ of the six measurements about the
weighted mean).
This is probably due to the considerable difficulties
in accounting for
the full measurement errors
(random and systematic) of astronomical
observations, but is nevertheless a source of concern,
particularly when viewed in conjunction with the
as yet poorly understood dispersion of D/H values
in the Milky Way (Linsky et al. 2006).

Additional measurements of the D/H ratio
at high redshift should, in principle at least, lead to
a more precise estimate of (D/H)$_{\rm p}$,
as well as to a better assessment of the reasons for the
scatter of the existing values. Such improvements are
not only of interest in a cosmological context, but are
also relevant to Galactic chemical evolution models
in which the degree of astration of D is an important
diagnostic (e.g. Romano et al. 2006; Steigman, Romano \& Tosi 2007).
In this paper, we report observations of deuterium absorption
in a seventh high redshift absorber,
a very metal-poor damped Ly$\alpha$ system (DLA)
at $z_{\rm abs} = 2.61843$, and consider their implications
for the determination of (D/H)$_{\rm p}$, $\Obo h^2$, and
other cosmological parameters.\footnote{Damped Ly$\alpha$ systems
are a class of QSO absorbers defined by their high
column densities of neutral hydrogen,
$\log [N$(H\,{\sc i})/cm$^{-2}] \geq 20.3$;
see Wolfe, Gawiser, \& Prochaska (2005), and Pontzen et al. (2008) for
recent reviews of DLAs.}

\section{Observations}
\label{sec:obs}

The existence of a metal-poor DLA at $z_{\rm abs} = 2.61843$
in the line of sight to the bright ($V = 17.1$), $z_{\rm em} = 2.785$,
QSO Q0913+072 has been known for some time (Pettini et al. 1997;
Ledoux et al. 1998; Erni et al. 2006).
More recently, it has been realised that DLAs of low metallicity
are also likely to have simple kinematics, possibly reflecting
an underlying mass-metallicity relation of the host dark-matter
halos (Ledoux et al. 2006; Murphy et al. 2007; Prochaska et al. 2008; 
Pontzen et al. 2008).
The most metal-poor DLAs are thus likely to be among the most
promising candidates for follow-up high-resolution spectroscopy
aimed at resolving the isotope shift in high order lines of the Lyman
series.

To this end, we targeted Q0913+072 with a concerted series
of observations in 2007, using UVES (Dekker et al. 2000) on the ESO VLT2.
The data were acquired in service mode; including a few spectra
retrieved from the UVES data archive, the total exposure time
devoted to this QSO was 77\,550\,s.
Details of the data reduction process can be found in Pettini et al. (2008);
briefly, the final reduced and combined spectrum covers the wavelength
region from 3310\,\AA\ to 9280\,\AA\ with a resolution of 6.7\,km~s$^{-1}$
FWHM, sampled with $\sim 3$\,pixels. The signal-to-noise ratio
in the continuum
varies from ${\rm S/N} \simeq 35$ per pixel
at the redshifted wavelength
of the damped \lya\ line, to ${\rm S/N} \simeq 15$ just longwards
of the Lyman limit 
of the $z_{\rm abs} = 2.61843$ DLA.

Pettini et al. (2008) presented an analysis of the chemical composition
and kinematics of this and other metal-poor DLAs, from consideration
of metal absorption lines due to H, C, N, O, Al, Si, and Fe;  we
refer the interested reader to that paper for details.
Of relevance to the present work are the following main results.
With relative abundances [C/H]\,$= -2.75$ and [Fe/H]\,$= -2.80$
the $z_{\rm abs} = 2.61843$ system in Q0913+072 is the most metal-poor
DLA known.\footnote{With the usual definition,
[X/H]\,$\equiv \log($X/H)$_{\rm DLA} -  \log($X/H)$_{\odot}$.}
Its oxygen abundance, [O/H]\,$= - 2.37$, is also among the lowest
recorded in damped systems.
The metal lines arising in the neutral gas have a very simple
kinematic structure, consisting of two components at
$z_{\rm abs} = 2.61828$ and $z_{\rm abs} = 2.61843$
($\Delta v = 12.4$\,km~s$^{-1}$),
with low internal velocity dispersions, $b = 3.7$\,km~s$^{-1}$
and 5.4\,km~s$^{-1}$ respectively (where $b = \sqrt{2} \sigma$
and $\sigma$ is the one dimensional velocity dispersion
of the absorbing atoms along the line of sight).
The column density of neutral gas is distributed between
the two components in approximately 3:7 proportion
(see Figure~1 and Table~2 of Pettini et al. 2008).

While these characteristics bode well for the possibility of resolving
the D component of the Lyman series lines, they do not guarantee it.
It is often the case that partially ionised gas at nearby velocities,
with optical depths too low to be recognised in the metal lines,
can blend with the H and D absorption due to the DLA itself,
and prevent a reliable measurement of $N$(D\,{\sc i}).
Partially ionised gas does in fact appear to contribute to
the C\,{\sc ii} absorption lines in this DLA, but seems to be confined to
positive, rather than negative, velocities relative to
$z_{\rm abs} = 2.61843$.
As can be seen from Fig.~\ref{fig:Q0913_montage},
this is indeed, somewhat fortuitously, the case.

\section{Hydrogen and Deuterium Absorption in the $z_{\rm abs} = 2.61843$ DLA}
\label{sec:HandD_inDLA}

\subsection{Deuterium}

\begin{figure*}
 \vspace*{-1cm}
  \centering
  {\includegraphics[angle=0,width=165mm]{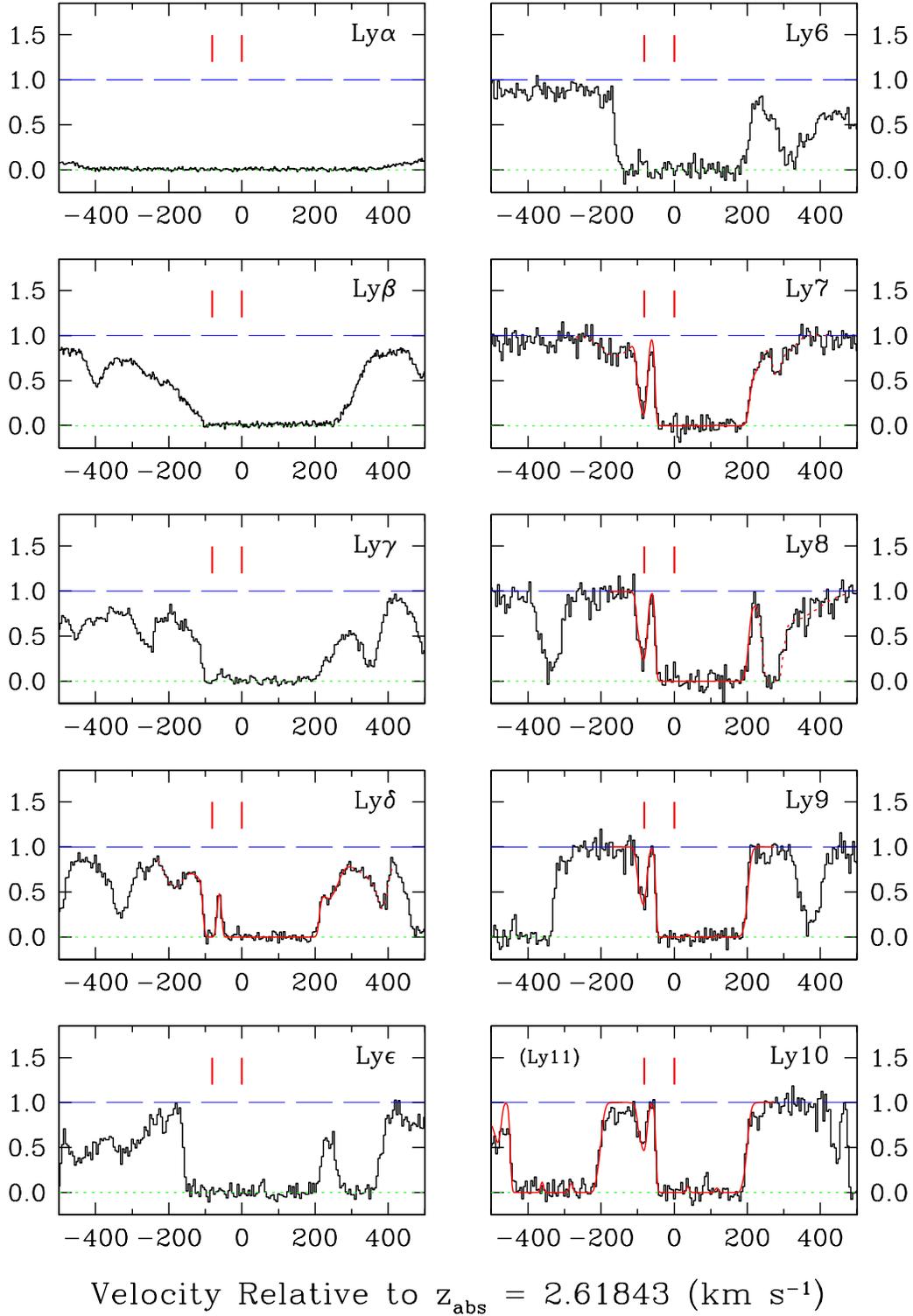}}
  \caption{
   Observed profiles (black histograms) and fitted Voigt profiles
   (continuous red lines) of absorption lines in the Lyman series of
   the $z_{\rm abs} = 2.61843$ DLA in Q0913+072.
   The $y$-axes of the plots show relative
   intensity.
   The two vertical tick marks in each panel indicate the expected
   locations of the main absorption component of the DLA
   in, respectively, H\,{\sc i} (at $v = 0$\,km~s$^{-1}$
   in the plots) and D\,{\sc i} (at $v = -81.6$\,km~s$^{-1}$).
   A second absorption component, centred at $v = -12.4$\,km~s$^{-1}$,
   is resolved in the metal lines associated with this DLA, but
   the two components are blended in the intrinsically broader
   H\,{\sc i} and D\,{\sc i} absorption. Additional H\,{\sc i}
   absorption is seen at positive velocities relative to the DLA,
   but its column density is $\sim 100$ times lower than that
   of the DLA, and therefore does not contribute to the
   observed D\,{\sc i} absorption lines (see text).
   }
   \label{fig:Q0913_montage}
\end{figure*}

In Fig.~\ref{fig:Q0913_montage} we have reproduced on a velocity
scale relative to $z_{\rm abs} = 2.61843$ the normalised profiles
of absorption lines in the Lyman series of the DLA, from \lya\
to Ly10. Ly11 is also visible in the Ly10 panel, while higher order
lines are all blended with one another, constituting an `effective'
Lyman limit from $\lambda_0 = 917$\,\AA.
It is clear from the high order transitions that there is
H\,{\sc i} absorption at positive velocities relative
the $z_{\rm abs} = 2.61843$ DLA, extending to $v \simeq +200$\,km~s$^{-1}$.
On the blue side, however, there appears to be no additional
components over those seen in the metal lines, and the D\,{\sc i}
absorption is clearly resolved in Ly$\delta$, and Ly7 through to Ly11,
six transitions in total.
(In Ly$\alpha$, Ly$\beta$ and Ly$\gamma$, which have the
highest transition probabilities, H\,{\sc i} and D\,{\sc i} are
intrinsically blended, whereas in Ly$\epsilon$ and Ly6 presumably
unrelated absorption in the Lyman forest is blended with the
absorption due to the DLA).

The availability of many D\,{\sc i} transitions
with a range of optical depths, from the saturated
Ly$\delta$ to the optically thin Ly10 and Ly11,
allows for a precise determination of the column
density $N$(D\,{\sc i}).
To this end, we fitted the profiles of
the Lyman lines where the D\,{\sc i} absorption
is resolved with theoretical
Voigt profiles generated by the VPFIT (version 8.02)
software package.\footnote{VPFIT is available from
http://www.ast.cam.ac.uk/\textasciitilde rfc/vpfit.html}
VPFIT uses $\chi^2$ minimisation to
deduce the values of redshift $z$,
logarithmic column density $\log [N$/(cm$^{-2}$)],
and Doppler parameter $b$ (km~s$^{-1}$)
that best reproduce the observed absorption line profiles,
taking into account the
instrumental broadening function
in its $\chi^2$ minimisation and error
evaluation.
We used the compilation of laboratory
wavelengths $\lambda_{\rm lab}$
and $f$-values by Morton (2003).
In Table~\ref{tab:Q0913_cloudmodel} we have collected relevant details
of the absorption model fitted to the observed line profiles in Fig.~\ref{fig:Q0913_montage};
the corresponding theoretical  profiles generated by VPFIT are shown with continuous
red lines in Fig.~\ref{fig:Q0913_montage}.

\begin{table}
\centering
    \caption{{Absorption components fitted to the
$z_{\rm abs} = 2.61843$ DLA in Q0913+072.}}
    \begin{tabular}{@{}llll}
    \hline
    \multicolumn{1}{c}{Ion}
& \multicolumn{1}{c}{$z_{\rm abs}$}
& \multicolumn{1}{c}{$b$}
& \multicolumn{1}{c}{Fraction}\\
    \multicolumn{1}{c}{}
& \multicolumn{1}{c}{}
& \multicolumn{1}{c}{(km~s$^{-1}$)}
& \multicolumn{1}{c}{}\\
    \hline
H\,{\sc i}$^{\rm a}$ & 2.61828$^{\rm b}$    & 11.9$^{\rm c}$  &  0.33$^{\rm d}$ \\
H\,{\sc i}$^{\rm a}$ & 2.61843$^{\rm b}$    & 11.1$^{\rm c}$  &  0.67$^{\rm d}$ \\
D\,{\sc i} & 2.61828$^{\rm b}$    & 8.6  &  0.33 \\
D\,{\sc i} & 2.61843$^{\rm b}$    & 8.7  &  0.67 \\
O\,{\sc i} & 2.61828    & 3.5  &  0.31 \\
O\,{\sc i} & 2.61843    & 5.2  &  0.69 \\
    \hline
    \end{tabular}

\flushleft{\hspace{1.3cm}$^{\rm a}${Additional components at positive velocities relative}\\
\hspace{1.5cm}{to $z_{\rm abs} = 2.61843$ were fitted to the H\,{\sc i} lines.}\\
\hspace{1.3cm}$^{\rm b}${Fixed to be the same as O\,{\sc i}.}\\
\hspace{1.3cm}$^{\rm c}${Poorly determined because all H\,{\sc i} lines are}\\
\hspace{1.5cm}{saturated and blended.}\\
\hspace{1.3cm}$^{\rm d}${Fixed to be the same as D\,{\sc i}.}}
    \label{tab:Q0913_cloudmodel}
\end{table}

As mentioned in Section 2, the metal lines detected in this DLA
consist of two components (Pettini et al. 2008),
with the parameters listed in Table~\ref{tab:Q0913_cloudmodel}
for O\,{\sc i}. The two components are not clearly distinguished
in D\,{\sc i} because: (i) absorption by the lighter D is intrinsically
broader (see discussion below),  and (ii) the S/N ratio of the
spectrum is lower at the shorter wavelengths of the high
order Lyman lines (at $z_{\rm abs} = 2.61843$
the wavelength region between Ly$\delta$ and the Lyman limit
is redshifted to $\lambda_{\rm obs} = 3438$--3318\,\AA,
approaching the cut-off by atmospheric ozone).
Thus, in the first step of the fitting process, we fixed the redshifts of the two
components to be the same as those determined from the metal lines
while keeping the values of $b$ and $N$(D\,{\sc i}) as free parameters
to be optimised by VPFIT.
The best fitting values of  $b$ and $N$(D\,{\sc i}) thus reached
are consistent with those deduced for O\,{\sc i} and other
metal lines. The relative
proportions of atoms between the two components are similar
in all species and the $b$-values are broader in D than in O,
as expected. Recall that the velocity dispersion of an
absorption line is the quadratic combination of macroscopic
(turbulence) and microscopic (temperature related) terms,
i.e. $b_{\rm tot}^2 = b_{\rm turb}^2 + b_{\rm T}^2$;
while the former is presumably  the same for all elements,
the latter has an inverse dependence on the square root of the
mass, since $b_{\rm T}^2 = 2kT/m$ (e.g. Str{\"o}mgren 1948).
The higher $b$-values of the two absorption components
in D than O imply temperatures of a few
thousand degrees.

The model parameters listed in Table~\ref{tab:Q0913_cloudmodel}
correspond to a best fitting value for the column density
of D\,{\sc i} (sum of the two components) of
\begin{center}
$\log [N$(D\,{\sc i})/cm$^{-2}] = 15.78\pm 0.02$,
\end{center}
but we
stress than this value does \emph{not} depend on the details of
the `cloud' model adopted. To test its robustness,
we rerun VPFIT without  any assumptions
about the distribution of velocities of the absorbers.
Using as a starting point a
single absorbing component with unspecified redshift
and $b$-value, VPFIT returned a best-fitting value,
$\log [N$(D\,{\sc i})/cm$^{-2}] = 15.79\pm 0.02$,
that is only 0.01\,dex higher than that obtained
with the parameters in Table~\ref{tab:Q0913_cloudmodel}
(although this second model results in a higher value of $\chi^2$
between computed and observed profiles).
Any other adjustment to the details of the profile
fits, such as varying the relative proportions of D\,{\sc i}
between the two components and their $b$-values,
or changing the parameters of nearby absorption in the
Lyman forest, resulted in even smaller differences in
the total value of $\log N$(D\,{\sc i}) returned by VPFIT.

As a final comment here, we point out that the VPFIT-generated
line profiles shown
in Figure~\ref{fig:Q0913_montage}
include a number of other components to the red of
$z_{\rm abs} = 2.61843$, required to
reproduce the saturated profiles of the H\,{\sc i} Lyman
lines which extend to $v \simeq 200$\,km~s$^{-1}$.
The parameters of these extra absorptions are poorly
constrained, but do not influence the determination
of $\log N$(D\,{\sc i}) because they involve column
densities of less than $\sim 1/100$ that of the DLA and thus
do not contribute to the observed D\,{\sc i} absorption.

\subsubsection{Consistency check with the apparent optical depth method}
Since the D\,{\sc i} absorption lines we observe are resolved,
we can deduce the column density directly from the measured
residual intensity in each wavelength, or velocity, interval across
the absorption lines (e.g. Hobbs 1974).
The apparent optical depth at velocity $v$, $\tau_a(v)$, 
is related to the observed intensity
in the line, $I_{\rm obs}(v)$, by
\begin{equation}
\tau_a(v) ~=~ -\ln~ [I_{\rm obs}(v)/I_0(v)] ,
\label{eq:tau}
\end{equation}
where $I_0(v)$ is the intensity in the continuum.
With the assumption of negligible smearing of the 
intrinsic line profile by the 
instrumental broadening function, we have:
\begin{equation}
\tau(v) \approx \tau_a(v) .
\end{equation}
The optical depth $\tau(v)$ is in turn related to the column density
of D\,{\sc i} atoms in each velocity bin,  
$N (v)$ in units of cm$^{-2}$~(km~s$^{-1}$)$^{-1}$, 
by the expression
\begin{equation}
N (v) ~=~ \frac{\tau(v)}{f\lambda} \times \frac{m_e c}{\pi e^2} 
~=~  \frac{\tau(v)}{f\lambda{\rm (\AA)}} \times 3.768 \times 10^{14}
\label{eq:N_optical_depth1}
\end{equation}
where the symbols $f$, $\lambda$, $c$, $e$ and $m_e$ have their usual
meanings.

As emphasized by Savage \& Sembach (1991),
the attraction of the apparent optical depth method
lies in the fact that no assumption has to be made
concerning the velocity distribution of the absorbers.
Furthermore, this method provides a consistency check when 
two or more transitions arising from the same atomic energy level,
but with different values of the product $f \lambda$, 
are analyzed, as is the case here.
The run of $N(v)$ with $v$ should be the same, within the errors
in $I_{\rm obs}(v)$, for all such lines.

For D\,{\sc i} in Q0913+072, we can apply to apparent optical
depth analysis to four unsaturated transitions, from Ly~7 to Ly~10;
the other lines in the D\,{\sc i} Lyman series being either partly
blended or saturated (see Figure~\ref{fig:Q0913_montage}).
For each transition, we deduced a value of $N$(D\,{\sc i}) 
by summing eq.~(\ref{eq:N_optical_depth1}) over the $n$ velocity
bins which make up the absorption profile:
\bneq
N_{\rm TOT} = \sum_{i=1}^{n} N_i(v).
 \label{eq:N_optical_depth2}
\eneq
From the known error in the value of $I_{\rm obs}(v)$
in each velocity bin, $\delta I_{\rm obs}(v)$,
we calculated the error $\delta N_{\rm TOT}$:
\bneq
\delta N_{\rm TOT}^2 = \sum_{i=1}^{n} \delta N_i(v)^2,
 \label{eq:N_optical_depth_error}
\eneq
which is asymmetric about $N_{\rm TOT}$ because of the
non-linear nature of eq.~(\ref{eq:tau}).

\begin{table}
\centering
    \caption{Results of apparent optical depth analysis of D\,{\sc i} lines}
    \begin{tabular}{@{}llll}
    \hline
    \multicolumn{1}{c}{Transition}
& \multicolumn{1}{c}{$\lambda_{\rm lab}^{\rm a}$}
& \multicolumn{1}{c}{$f^{\rm a}$}
& \multicolumn{1}{c}{$N_{\rm TOT}$(D\,{\sc i})}\\
    \multicolumn{1}{c}{}
& \multicolumn{1}{c}{(\AA)}
& \multicolumn{1}{c}{}
& \multicolumn{1}{c}{($10^{15}$\,cm$^{-2}$)}\\
    \hline
\smallskip
D\,{\sc i}\, Ly7 & 925.9737    & 0.003184 &  $5.64^{+0.43}_{-0.25}$\\
\smallskip
D\,{\sc i}\, Ly8 & 922.899    & 0.002216 &  $6.35^{+0.40}_{-0.29}$\\
\smallskip
D\,{\sc i}\, Ly9 & 920.712    & 0.001605 &  $6.08^{+0.37}_{-0.32}$\\
\smallskip
D\,{\sc i}\, Ly10 & 919.102    & 0.001201 &  $5.61^{+0.41}_{-0.37}$\\
\\
Weighted Mean &  & & $5.93^{+0.21}_{-0.16}$\\
    \hline
    \end{tabular}

\flushleft{\hspace{0.53cm}$^{\rm a}${Morton (2003).}
}
    \label{tab:N_aod}
\end{table}

\begin{figure}
  \centering
  {\hspace*{-0.3cm}\includegraphics[angle=0,width=86mm]{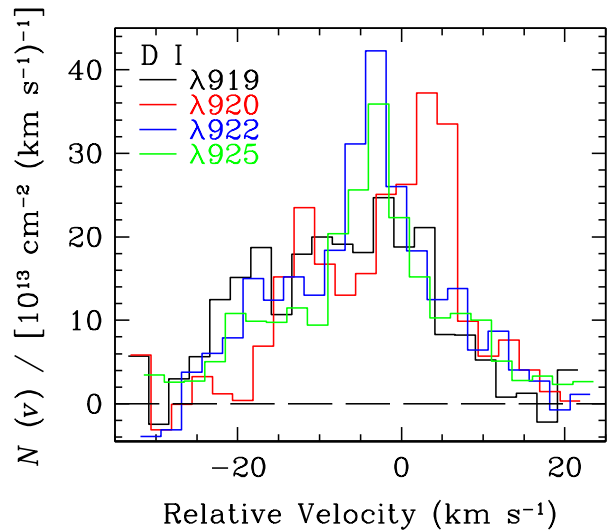}}
\vspace{-0.5cm}
  \caption{
   Apparent optical depth analysis: column density as a function of velocity 
   for the four unsaturated and unblended
   D\,{\sc i} lines in the $z_{\rm abs} = 2.61843$ DLA in Q0913+072.
   The transitions shown in the legend are in increasing order of 
   $f \lambda$ from the top. 
   Values of $N_{\rm TOT}$(D\,{\sc i}) are collected in Table~\ref{tab:N_aod}.
   }
   \label{fig:aod}
\end{figure}

Figure~\ref{fig:aod} shows the run of $N_i(v)$ across the 
four D\,{\sc i} absorption lines, while values of $N_{\rm TOT}$
and $\delta N_{\rm TOT}$ are listed in Table~\ref{tab:N_aod}.
As can be seen from Figure~\ref{fig:aod} and Table~\ref{tab:N_aod},
the four D\,{\sc i} absorption lines 
are in good mutual agreement, and the standard
deviation between the four independent measures of  $N$(D\,{\sc i})
is $1 \sigma = 0.35 \times 10^{15}$\,cm$^{-2}$, or $\sim 6\%$ of the mean.
The weighted mean of the four values of $N$(D\,{\sc i}) returned
by the apparent optical depth analysis
is $\langle N$(D\,{\sc i})$\rangle = (5.93^{+0.21}_{-0.16}) \times 10^{15}$\,cm$^{-2}$,
where the uncertainties quoted are the errors on the weighted mean. 
On a logarithmic scale $\log [N$(D\,{\sc i})/cm$^{-2}] = 15.775 \pm 0.014$, in
very good agreement (as expected) with the value of 
$\log [N$(D\,{\sc i})/cm$^{-2}] = 15.78 \pm 0.02$ returned by VPFIT,
which we retain in the subsequent analysis.

\subsection{Hydrogen}

\begin{figure*}
 \vspace*{-0.5cm}
  \centering
  {\hspace*{0.4cm}\includegraphics[angle=270,width=175mm]{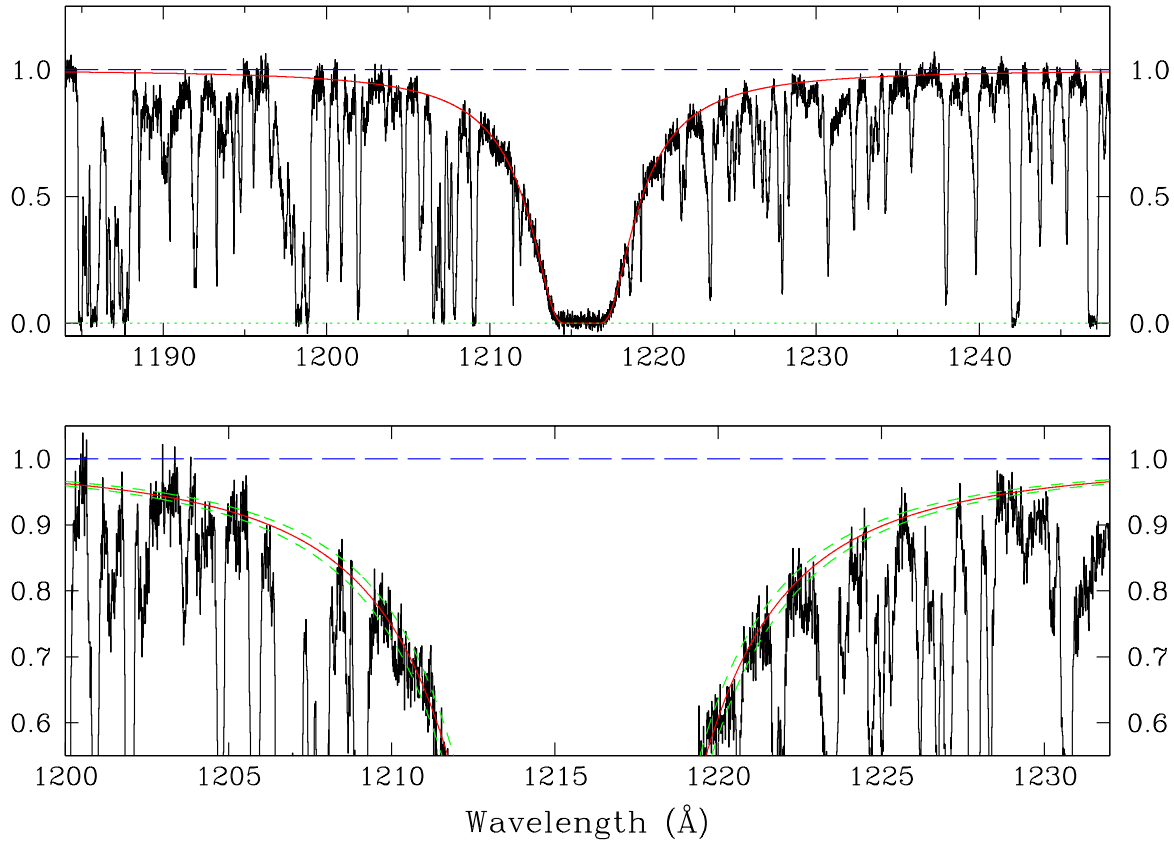}}
 \vspace{-0.75cm}
  \caption{
   \textit{Top:} Portion of the normalised spectrum of Q0913+072
   encompassing the damped \lya\ line at
   $z_{\rm abs} = 2.61843$ (black histogram).
    The continuous red line shows the theoretical
   profile for a neutral hydrogen column density
   $\log [N$(H\,{\sc i})/cm$^{-2}] = 20.34$.
   \textit{Bottom:} Expanded portion of the upper plot on a scale
   chosen to illustrate the sensitivity of the damping wings to the 
   column density of neutral hydrogen. The short-dash green lines 
   correspond to the $\pm 0.04$ limits to the best fitting 
   value $\log [N$(H\,{\sc i})/cm$^{-2}] = 20.34$.
   In both panels the $y$-axis is residual intensity.
   }
   \label{fig:Q0913_lya}
\end{figure*}

Fig.~\ref{fig:Q0913_lya} shows the region encompassing
the damped \lya\ line. In principle, the column density
of neutral hydrogen should be tightly constrained from
the shape of the damping wings which extend
over many hundreds of pixels (and are thus extremely
well sampled by the data).
In practice, the limiting factors are the uncertainty in the
determination of the continuum level (to which the
absorption is normalised) and the overlapping absorption
from the narrower lines in the \lya\ forest.
After numerous trials varying the continuum level
and the weights given to different spectral intervals
deemed to be free of overlapping absorption
(see Kirkman et al. 2003 for a more extensive discussion
of the problem) we converged on the best fitting value:
\begin{center}
$\log [N$(H\,{\sc i})/cm$^{-2}] = 20.34\pm 0.04$;
\end{center}
the corresponding theoretical damped profile is
overplotted on the data in Fig.~\ref{fig:Q0913_lya}.

It is difficult to estimate reliably the systematic
error which may be affecting this determination.
While consistent with the above estimate of
$\log N$(H\,{\sc i}) from \lya,
the higher order Lyman lines do not help to reduce
the error further because the are all saturated and
their equivalent widths include uncertain contributions
from the lower column density components to the red
of the DLA.
In future, it may be possible to improve on the
accuracy with which values $N$(H\,{\sc i}) can
be deduced from the analysis of
damped profiles buried within the \lya\ forest
by developing more sophisticated statistical methods
to deal with the overlapping absorption, perhaps
analogous to those used to subtract foreground
Galactic emission from maps of the CMB.
For the moment, we can perhaps obtain an
indication of the magnitude of the systematic
uncertainty in $N$(H\,{\sc i}) by considering
three previous estimates of this quantity,
reported by different authors who used different
spectrographs and telescopes,
as follows:
$\log [N$(H\,{\sc i})/cm$^{-2}] = 20.36\pm 0.08$
(Pettini et al. 1997); $20.2 \pm 0.1$ (Ledoux et al. 1998);
$20.36 \pm 0.05$ (Erni et al. 2006).
Of these, only the last one refers to a (small)
subset of the data used here.
From these results it appears that our present estimate,
$\log [N$(H\,{\sc i})/cm$^{-2}] = 20.34\pm 0.04$,
is unlikely to be in error by more than the
stated margin.

From the values of $N$(D\,{\sc i}) and
$N$(H\,{\sc i}) deduced in sections 3.1 and 3.2,
we arrive at the determination of the deuterium
abundance in the $z_{\rm abs} = 2.61843$ DLA in
line to Q0913+072 of:
\begin{center}
$\log {\rm (D/H)} = -4.56 \pm 0.04$
\end{center}
(where the errors have been combined in quadrature).

\begin{table*}
\centering
\begin{minipage}[c]{0.7\textwidth}
    \caption{\textsc{Prime Sample of D/H measurements in QSO Absorption Line Systems}}
    \begin{tabular}{@{}lrrrrrr}
    \hline
   \multicolumn{1}{c}{QSO}
& \multicolumn{1}{c}{$z_{\rm em}$}
& \multicolumn{1}{c}{$z_{\rm abs}$}
& \multicolumn{1}{c}{$\log N$\/(H\,{\sc i})}
& \multicolumn{1}{c}{[O/H]$^{\rm b}$}
& \multicolumn{1}{c}{$\log {\rm (D/H)}$}
& \multicolumn{1}{c}{Ref.$^{\rm a}$}\\
    \multicolumn{1}{c}{}
& \multicolumn{1}{c}{}
& \multicolumn{1}{c}{}
& \multicolumn{1}{c}{(cm$^{-2}$)}
& \multicolumn{1}{c}{}
& \multicolumn{1}{c}{}
& \multicolumn{1}{c}{}\\
  \hline
HS\,0105+1619                      & 2.640     &  2.53600   & $19.42 \pm 0.01$   & $-1.70$                   &  $-4.60 \pm 0.04$    &  1 \\
Q0913+072                             & 2.785     &  2.61843   & $20.34 \pm 0.04$   & $-2.37$                   &  $-4.56 \pm 0.04$    &  2, 3 \\
Q1009+299                             & 2.640     &  2.50357   & $17.39 \pm 0.06$   & $< -0.67^{\rm c}$  &  $-4.40 \pm 0.07$    &  4 \\
Q1243+307                           & 2.558     &  2.52566   & $19.73 \pm 0.04$   & $-2.76$                   &  $-4.62 \pm 0.05$    &  5 \\
SDSS~J155810.16$-$003120.0  & 2.823     & 2.70262   & $20.67 \pm 0.05$    & $-1.47$                   & $-4.48 \pm 0.06$     &  6  \\
Q1937$-$101                          & 3.787     & 3.57220    & $17.86 \pm 0.02$   & $ < -0.9$                  & $-4.48 \pm 0.04$    & 7 \\
Q2206$-$199                          & 2.559     & 2.07624    & $20.43 \pm 0.04$   & $-2.04$                    & $-4.78 \pm 0.09$    & 2, 8 \\
   \hline
    \end{tabular}
    \smallskip

$^{\rm a}${References---1: O'Meara et al. (2001);
2: Pettini et al. (2008);
3: This work;
4: Burles \& Tytler (1998b);
5: Kirkman et al. (2003);
6: O'Meara et al. (2006);
7: Burles \& Tytler (1998a);
8: Pettini \& Bowen (2001).
}\\
$^{\rm b}${Relative to the solar value $\log ({\rm O/H})_{\odot} + 12 = 8.66$ (Asplund, Grevesse \& Sauval 2005).}\\
$^{\rm c}${This is a very conservative upper limit on the metallicity. Burles \& Tytler (1998b)
estimate [Si/H]\,$\simeq -2.5$ and [C/H]\,$\simeq -2.9$ from photoionisation modelling.
}\\
    \label{tab:DtoH}
\end{minipage}
\end{table*}

\section{The primordial abundance of deuterium}
\label{sec:D_abund_prim}

In Table~\ref{tab:DtoH} we have collected relevant measurements
for all the high redshift QSO absorption systems
where the isotope shift has been resolved
in absorption lines of the Lyman series.
This prime sample of what are generally considered
to be the most reliable measures of D/H at high $z$
now consists of seven independent determinations.
Other reports in the literature (e.g. Levshakov et al. 2002;
Chrighton et al. 2004), while still interesting, refer to
spectra of less straightforward interpretation because not all
the D\,{\sc i} components are resolved, so that
the values of D/H deduced are more dependent on
the precise description of the kinematics of the gas
than is the case for the seven absorbers listed in Table~\ref{tab:DtoH}
(see also the discussions of this point by Kirkman et al. 2003
and O'Meara et al. 2006).

All the measurements of D/H in Table~\ref{tab:DtoH}
should be representative
of the primordial abundance (D/H)$_{\rm p}$, because
in all seven cases the gas has undergone
little chemical enrichment, as evidenced by the low
abundance of O (and other heavy elements),  less
than 1/10 of solar (see column 5 of Table~\ref{tab:DtoH}).
For comparison, the total degree of astration of D over the
lifetime of the
Milky Way amounts to only
$\simlt 20$\%,  if one accepts the possibility
that in the local interstellar medium some
of the deuterium may be depleted onto dust grains
(Linsky et al. 2006).
Theoretically, galactic chemical evolution models
(e.g. Prantzos \& Ishimaru 2001; Romano et al. 2006)
show negligible reduction in D/H from the primordial
value when the gas metallicities are as low as those of
the seven QSO absorbers in Table~\ref{tab:DtoH},
while observationally no trend is observed between
D/H and O/H (Pettini 2006). Dust depletion of D
is not expected to be an issue here, given the very low
depletions of even the most refractory elements
at these low  levels of chemical enrichment (e.g. Akerman et al. 2005).

We arrive at an estimate of $\log {\rm (D/H)_p}$ by averaging
the individual measure of $\log {\rm (D/H)}$ in Table~\ref{tab:DtoH}
to obtain the weighted mean
$\langle \log {\rm (D/H)_p}\rangle = -4.55 \pm 0.02$.
However, the scatter of the points about this value is
rather high for the quoted error bars assuming a Gaussian error model,
giving $\chi^2 = 19$ which formally corresponds to
a high probability that a further independent experiment would
obtain a better fit to to the data, $P(\chi^2 > 19) < 0.01$.
This suggests that the errors on the individual
measures of  D/H may have been underestimated --
a well known and much discussed (e.g. Steigman 2007)
`problem' of the determination of (D/H)$_{\rm p}$
from QSO spectra.
An alternative, and more conservative, method of
estimating the error on the weighted mean
would attempt to account for the actual scatter in the data.
One possibility
is to consider the mean and
standard deviation of a large number of weighted means
generated by random sampling (with substitution) of the seven observed
values (the bootstrap method; see Efron \& Tibshirani 1993).
In this way we obtain:
\begin{center}
$\langle \log {\rm (D/H)_p} \rangle = -4.55 \pm 0.03$
\end{center}
A more Bayesian approach could try to account
for uncertainty in the error bars
using the observed scatter in the data.
For example, fitting a Gaussian model
with the variance on each point increased
by a constant to $\sigma_i^2 + \delta\sigma^2$ shows that
the data prefer values of $\delta\sigma > 0$.
Using the maximum likelihood value $\delta\sigma\sim 0.07$
gives a posterior constraint on $\langle \log {\rm (D/H)_p} \rangle$
consistent with the bootstrap estimate above,
as does marginalizing over $\delta\sigma^2$
with prior $\propto 1/(0.02^2+\delta\sigma^2)$
(though the distribution is somewhat non-Gaussian).
Using a model that multiplicatively increases the noise gives similar results.

\begin{figure}
\vspace*{-3.75cm}
  \centering
  {\hspace*{-1.2cm}\includegraphics[angle=0,width=106mm]{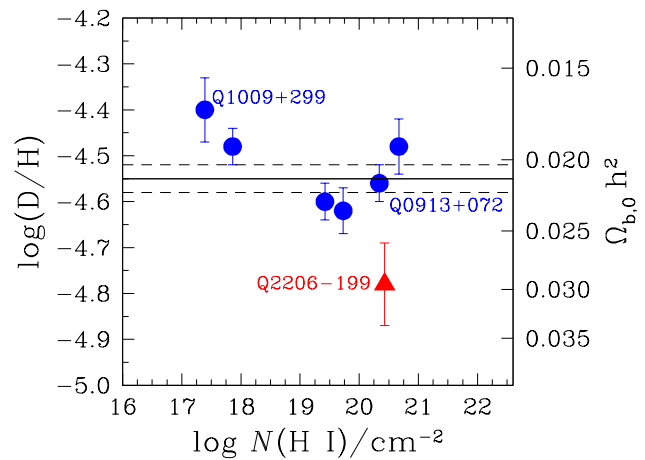}}
\vspace{-4.05cm}
  \caption{
   Measures of deuterium abundance in high redshift QSO absorbers.
   Only cases were the deuterium absorption is clearly resolved from
   nearby spectral features are shown here (see text). Blue circles
   denote systems observed from the ground with 8-10\,m telescopes
   and echelle spectrographs, while the red triangle refers to lower
   resolution observations made with the \textit{Hubble Space Telescope}.
   Absorption systems discussed in the text are labelled with the name
   of the background QSO. The horizontal lines are drawn at the
   weighted mean value of $\log {\rm (D/H)}$
   and its error, as determined with the bootstrap method.
   }
   \label{fig:DtoH}
\end{figure}

The scatter in the reported determinations of D/H in
QSO absorbers is illustrated in Figure~\ref{fig:DtoH}.
Out of the seven measurements, there are two which lie
outside the confidence intervals of the mean
by more than $1 \sigma$: Q1009+299 and Q2206$-$199.
If we consider these two cases more closely we can find,
in retrospect, plausible reasons why their $1 \sigma$ errors
may have been underestimated.
The partial Lyman limit system in line to Q1009+299 has the lowest column
density among the sample, $\log [N$(H\,{\sc i})/cm$^{-2}] = 17.39$,
and is unique among the ones considered here in showing
D\,{\sc i} absorption in only \emph{one} line, \lya,
the column density of D\,{\sc i} being too low
to produce discernible absorption in
higher order lines.
The $z_{\rm abs} = 2.07624$ DLA towards Q2206$-$199
is the lowest redshift absorber in the sample, requiring
space-borne observations to record the high order
lines where D\,{\sc i} absorption can be resolved from
H\,{\sc i}. The \textit{HST}-STIS spectrum of this object
published by Pettini \& Bowen (2001) is of lower S/N
ratio and resolution than the other six cases in Table~\ref{tab:DtoH}
which were all obtained with ground-based 8-10\,m telescopes
and echelle spectrographs.
If we arbitrarily double the $1 \sigma$ estimates of the errors
in the determinations of D/H in  Q1009+299 and Q2206$-$199
reported in the original works, we find the same weighted mean
$\langle \log {\rm (D/H)_p}\rangle = 4.55 \pm 0.02$ as before,
but a much reduced
$\chi^2 = 11$ ($P(\chi^2 > 11) \simeq  0.1$).

In conclusion, the value of D/H we deduce here for the
$z_{\rm abs} = 2.61843$ DLA in the spectrum of
Q0913+072 is in good agreement with four out of the
previous six determinations generally considered to be
the most reliable. Its inclusion in the `prime' sample,
helps identify two outliers and plausible reasons
for much of the scatter among the sample. The weighted mean
$\langle \log {\rm (D/H)_p}\rangle$ is unchanged
compared to the most recent previous estimate of this
quantity by O'Meara et al. (2006), leading us to
conclude that the true value of the primordial
abundance of deuterium lies in the range:
\bneq
\langle \log {\rm (D/H)_p} \rangle = -4.55 \pm 0.03
\label{eq:mean_dtoh}
\eneq
at the 68\% confidence level.

\subsection{The cosmological density of baryons from (D/H)$_{\rm p}$}

Steigman (2007) has pointed out that in the range of interest
here the primordial abundance of D is simply related via the
expression
\bneq
10^5 {\rm (D/H)_p} = 2.67 (1 \pm 0.03) \, (6/\eta_{10})^{1.6}
\label{eq:dtoh_vs_eta10}
\eneq
to $\eta_{10}$ which measures the universal ratio of the densities
of baryons and photons in units of $10^{-10}$:
\bneq
\eta_{10} \equiv 10^{10}({n_{\rm b}/}{n_\gamma}) = 273.9 \,\Omega_{\rm b,0} h^2
\label{eq:eta10_to_omega_b}
\eneq
The 3\% uncertainty in eq.~(\ref{eq:dtoh_vs_eta10}) is comparable
to the uncertainties in the nuclear reaction rates
used in Big-Bang nucleosynthesis codes.
The conversion from $\eta_{10}$ to $\Omega_{\rm b,0} h^2$
in eq.~(\ref{eq:eta10_to_omega_b}) is accurate to
about 0.1\% (Steigman 2006).
From eqs.~(6), (7) and (8), we find:
\bneq
\Omega_{\rm b,0}h^2{\rm  (BBN) } = 0.0213 \pm 0.0009 \pm 0.0004
\label{eq:omega_b(bbn)}
\eneq
where the error terms reflect the uncertainties in,
respectively,
(D/H)$_{\rm p}$ (eq.~\ref{eq:mean_dtoh}) and the
nuclear reaction rates (eq.~\ref{eq:dtoh_vs_eta10}).
Combining the two error terms in quadrature, we 
have:
\bneq
\Omega_{\rm b,0}h^2{\rm  (BBN) } = 0.0213 \pm 0.0010 
\label{eq:omega_b(bbn2)}
\eneq

\begin{figure*}
  \centering
  {\includegraphics[angle=0,width=95mm]{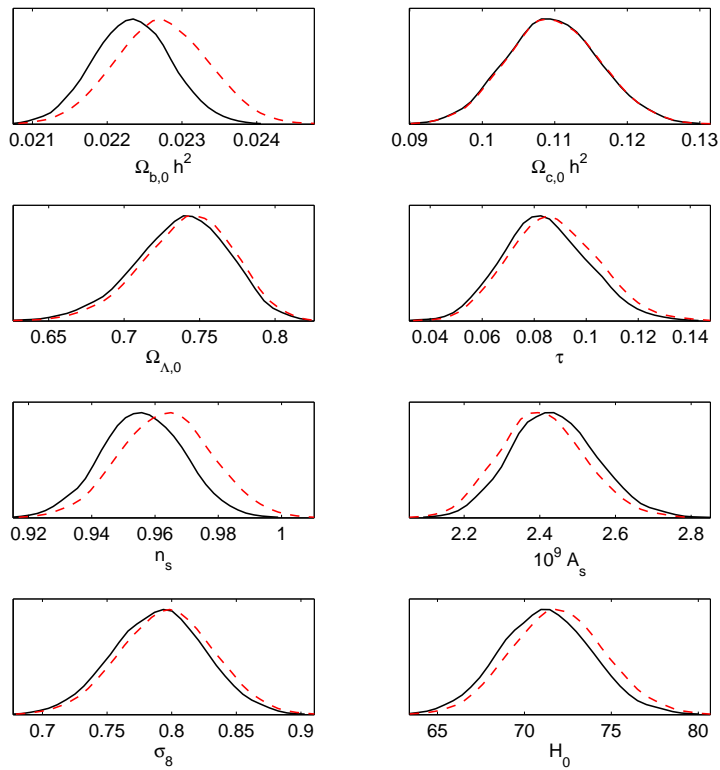}}
  \caption{
   Probability distributions of cosmological parameters deduced
   from the analysis of \textit{WMAP 5} data alone (red dashed line)
   and including $\Omega_{\rm b,0}h^2{\rm  (BBN) }$ as a prior
   (black continuous line).
    }
   \label{fig:cosm_para}
\end{figure*}

The analysis by Dunkley et al. (2008) of five years of observations
with the \textit{WMAP} satellite concluded that, on the basis of
the \textit{WMAP} data alone
\bneq
\Omega_{\rm b,0}h^2{\rm  (CMB) } = 0.02273 \pm 0.00062\,.
\label{eq:omega_b(cmb)}
\eneq
The two estimates of $\Omega_{\rm b,0}h^2$ agree (just) within the
errors; we also note that the uncertainty from BBN is now
comparable to that from the CMB. Given that the best-fitting
value of $\Omega_{\rm b,0}h^2{\rm  (CMB) }$ is tied to those
of other cosmological parameters, it is of interest to consider
the effect of including the value of
$\Omega_{\rm b,0}h^2{\rm  (BBN) }$ as a prior in the
analysis of the \textit{WMAP 5} data.

\section{Combined constraints on cosmological parameters
from the CMB and (D/H)$_{\rm p}$}
\label{sec:BBN_and_CMB}

The five-year \textit{WMAP} CMB temperature maps and
large scale polarization maps together provide tight constraints
on several combinations of cosmological parameters
(Hinshaw et al. 2008; Dunkley et al 2008).
In order to constrain \emph{individual} parameters, however,
it helps to apply external data that can break degeneracies.
The baryon density, which affects the relative heights of the CMB acoustic peaks,
is partly degenerate with the spectral index of primordial fluctuations, $n_{\rm s}$,
since \textit{WMAP} provides precision measurements of only two peaks in the temperature
power spectrum. Increasing the value of $n_{\rm s}$
increases the height of the second peak,
which can be compensated
by a decrease in the baryon density.
Hence we can get a better constraint on the spectral index
by combining the CMB data with an independent determination
of  $\Omega_{\rm b,0} h^2$.

Among the alternative avenues to $\Omega_{\rm b,0}$ which have been considered,
the primordial abundance of deuterium is the most accurate at present.
The baryonic acoustic oscillations imprinted
in the large-scale distribution of galaxies currently
provide less stringent constraints on $\Omega_{\rm b,0}$
than either the CMB or (D/H)$_{\rm p}$
and, in any case, really measure the combination of
parameters $\Obo /\Omo h$ (e.g. Blake et al. 2007).
Measurements of other light elements
created in Big-Bang nucleosynthesis
suffer from systematic uncertainties which are difficult to quantify
(in the case of helium) or are poorly understood (for lithium), 
as discussed in recent reviews by Molaro (2008) and Steigman (2007)
(see also Simha \& Steigman 2008). 
In considering only deuterium in our joint analysis
with the CMB we make the implicit assumption that
at present the primordial abundances of  $^4$He and $^7$Li
are difficult to reconcile with that of D because of
astrophysical considerations (such as how to extrapolate
from measured values in local astrophysical environments
to the primordial abundances) and do not reflect
an underlying departure from standard Big-Bang nucleosynthesis.
If the latter were the case, we would be unjustified in
comparing the CMB fluctuations with only (D/H)$_{\rm p}$.

Parameter constraints from CMB data are usually encoded
in a set of samples from the posterior
distribution generated from the likelihood function
by Markov Chain Monte Carlo methods.
In combining the CMB data with the value reached here from
(D/H)$_{\rm p}$, $\Omega_{\rm b, 0} h^2  {\rm (BBN)} =  0.0213 \pm 0.0010$,
we assume that the baryon density likelihood is an uncorrelated Gaussian.
Since the constraint is only on one parameter,
and consistent with the baryon density inferred from the CMB alone,
we can use importance sampling to re-weight the parameter samples
with the extra constraint (see Lewis \& Bridle 2002 for details).
For each sample in the original chain supplied
by the WMAP team,\footnote{Available from: \\
http://lambda.gsfc.nasa.gov/product/map/dr3/parameters.cfm}
we weight the sample by
$\exp\left[-(\Omega_{\rm b,0} h^2 - 0.0213)^2/(2\times 0.001^2)\right]$.
Marginalized constraints for individual parameters
can be calculated easily from the weighted chains;
we use the `GetDist' program provided with
CosmoMC   (Lewis \& Bridle 2002) to do this.

\begin{figure}
  \centering
  {\includegraphics[angle=0,width=85mm]{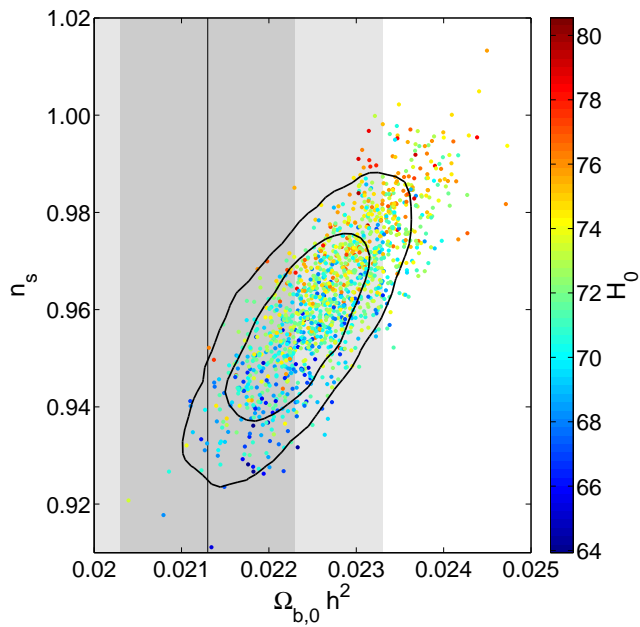}}
  \caption{
   The points show pairs of values of the baryon density, $\Omega_{\rm b,0}h^2$, and 
   the spectral index of primordial fluctuations, $n_{\rm s}$, implied by the 
   \textit{WMAP} data alone, colour-coded according to their value of the
   Hubble parameter $H_0$ (in units of km~s$^{-1}$~Mpc$^{-1}$).  
   The vertical line shows the value of $\Omega_{\rm b,0}h^2{\rm  (BBN) }$,
   with the shaded regions showing the $1 \sigma$ and $2 \sigma$ bounds
   from eq.~(\ref{eq:omega_b(bbn2)}). 
   The contours are the $1 \sigma$ and $2 \sigma$ constraints obtained by
   combining the \textit{WMAP} data with (D/H)$_{\rm p}$.
   }
   \label{fig:n_vs_Omega_b}
\end{figure}

\subsection{Standard $\Lambda$CDM cosmological model}
\label{sec:LCDM}

\begin{figure*}
  \centering
  {\includegraphics[angle=0,width=95mm]{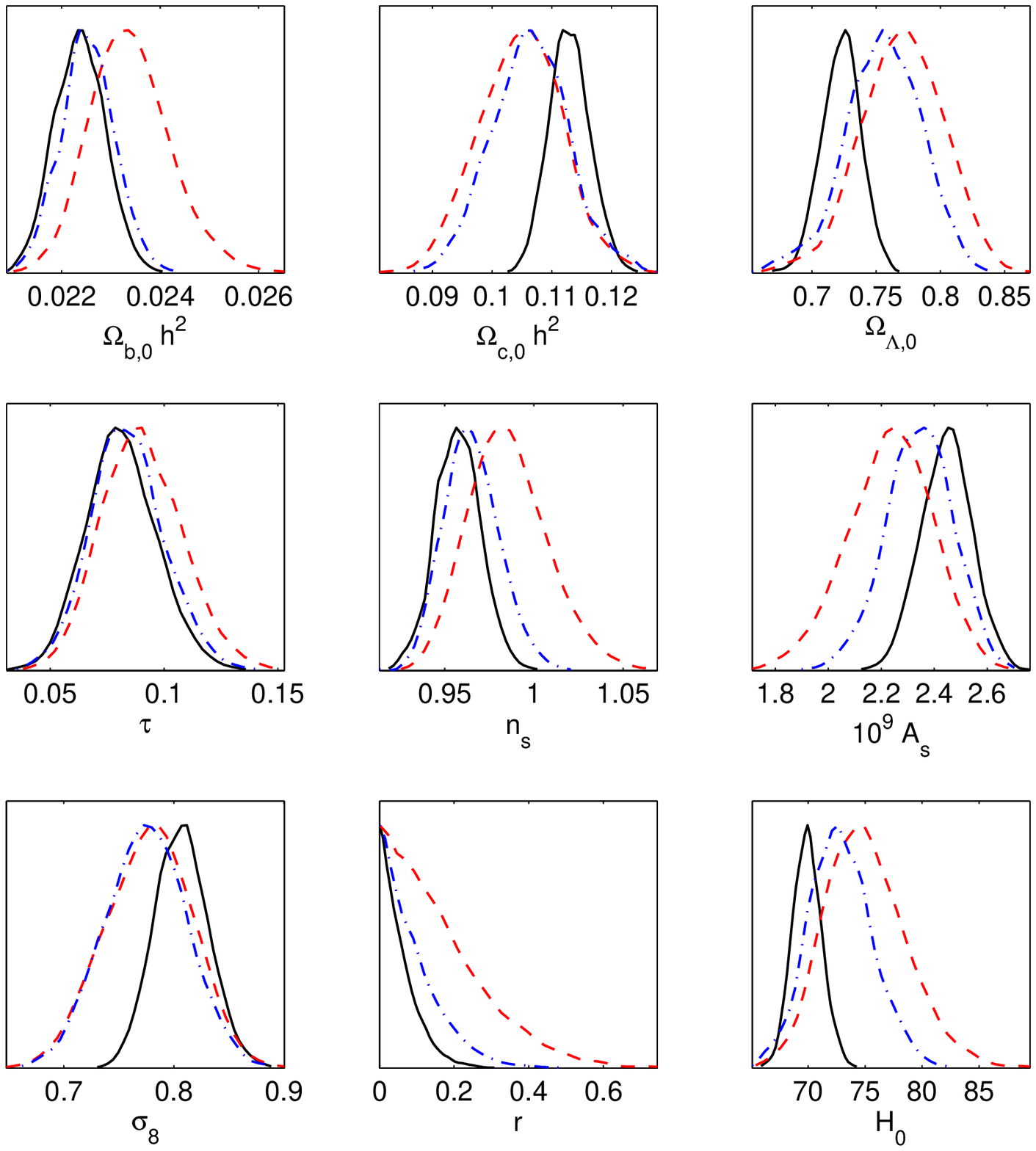}}
  \caption{
   Probability distributions of cosmological parameters deduced
   from the analysis of: (i) \textit{WMAP~5} data alone (red dashed line);
   (ii) \textit{WMAP~5} + (D/H)$_{\rm p}$ (blue dot-dash line);
   (iii) adding to (ii) the constraints imposed by baryon acoustic oscillations
   in the large-scale distribution of galaxies and by
   the distances to Type Ia supernovae
   (black continuous line -- see text for relevant references).
     }
   \label{fig:cosm_para_tensors}
\end{figure*}

We assume the simplest flat $\Lambda$CDM cosmological model,
with a power-law purely adiabatic spectrum 
of linear primordial curvature perturbations
with spectral index $n_{\rm s}$ and 
amplitude $A_{\rm s}$ at $k=0.002\,{\rm Mpc}^{-1}$,
dark matter density $\Omega_{\rm c, 0} h^2$,
baryon density $\Omega_{\rm b, 0} h^2$,
optical depth for sharp reionization $\tau$, and
cosmological constant density relative to 
critical $\Omega_{\Lambda, 0}$ (with a flat prior).
Marginalized one-dimensional parameter constraints obtained in this
way are shown in Fig.~\ref{fig:cosm_para}.

The combined constraint on the baryon density is
$\Omega_{\rm b,0} h^2 = 0.0223 \pm 0.0005$
where the error corresponds to the 68\% probability
from the distribution shown in the top left-hand panel
of Fig.~\ref{fig:cosm_para}.
For comparison, the value obtained by
combining the \textit{WMAP~5} data
with the distance measurements from Type Ia supernovae
and the baryon acoustic oscillations in the large-scale
distribution of galaxies is
$\Omega_{\rm b,0} h^2 = 0.02265 \pm 0.0006$
(Komatsu et al. 2008).

Since the (D/H)$_{\rm p}$ measurement prefers values of the baryon density
towards the lower end of the range allowed by the CMB,
inclusion of this prior leads to a lower value of the spectral index:
we obtain $n_{\rm s} =0.956\pm 0.013$, and
$n_{\rm s}< 0.990$ at 99\% confidence (purely statistical errors).
Thus, red spectral tilts are preferred and
a Harrison-Zeldovich spectrum with $n_{\rm s}=1$
is ruled out in the simplest $\Lambda$CDM models.
Other parameter constraints are almost unchanged,
with only slight shifts in $\tau$, $\sigma_8$ and $H_0$ towards lower values.
The combined goodness-of-fit parameter $\chi^2_{\rm eff}$
increases by about one on including the $\Obo h^2 {\rm (BBN)}$ constraint,
consistent with expectations for adding one
independent parameter.
Figure~\ref{fig:n_vs_Omega_b} shows the combined constraints
on $n_{\rm s}$ and $\Omega_{\rm b,0} h^2$ with and without 
inclusion of the (D/H)$_{\rm p}$ prior.

\subsection{Constraints on a tensor component to primordial fluctuations}
\label{sec:BBN_and_CMB_tensors}

If the spectral index is less than one, some inflationary models predict an
observable amplitude of primordial gravitational waves. Adding a parameter
$r$ (with flat prior) to measure the ratio of the tensor and scalar power at
$k=0.002\,{\rm Mpc}^{-1}$, is therefore well motivated
(using the relation between
the tensor and scalar tilts for the simplest inflation
models; see Komatsu et al. 2008).
A combination of \textit{WMAP~5} data and the (D/H)$_{\rm p}$
constraint then gives $n_{\rm s} < 1.00$, and a limit on the tensor amplitude
$r < 0.26$, both at 95\% confidence.
The constraints on these parameters are
consistent with, and comparable to, those obtained by Komatsu et al. (2008)
by considering \textit{WMAP} data in conjunction with
baryon oscillations and supernova distances measurements,
but have completely different systematics.
Considering all the constraints together -- \textit{WMAP~5},
(D/H)$_{\rm p}$, Type Ia supernovae (Riess et al. 2004; Astier et al. 2006;
Wood-Vasey et al. 2007), and baryon acoustic oscillations in the large-scale
distribution of galaxies (Percival et al. 2007 ) -- we arrive at the full joint
probability distributions shown by the black continuous lines
in Fig.~\ref{fig:cosm_para_tensors}.
We find $r < 0.16$ at 95\% confidence and
$n_{\rm s} < 0.994$ at 99\% confidence.
These limits strongly constrain possible
models of the early universe, but are consistent with
some of the simplest
inflationary (and other) models. Joint marginalized constraints on the
various parameters are:
$\Omega_{\rm b,0}h^2 = 0.0224 \pm  0.0005$,
$\Omega_{\rm c, 0} h^2 = 0.1130 \pm 0.0034$,
$\Omega_{\Lambda, 0} = 0.723\pm 0.015$,
$h = 0.700\pm 0.013$,
$n_{\rm s} = 0.959 \pm 0.013$, $\sigma_8 = 0.808 \pm 0.025$,
and $\tau = 0.818 \pm 0.016$  ($1 \sigma$ errors).

\subsection{A closer look at the impact of the errors
on D/H on the derivation of cosmological parameters}

The results presented in 
Sections~\ref{sec:LCDM} and 
\ref{sec:BBN_and_CMB_tensors}
assumed a Gaussian error on 
$\Omega_{\rm b,0} h^2$(BBN) derived
from the dispersion of D/H measurements 
analysed with the bootstrap method.
However, values for high-significance confidence limits are 
quite sensitive to the shape of the tails of the distribution
and it is therefore
worth assessing the robustness of our results to changes 
in the statistical model. 

\begin{figure}
  \centering
 {\hspace*{-0.35cm} \includegraphics[angle=0,width=80mm]{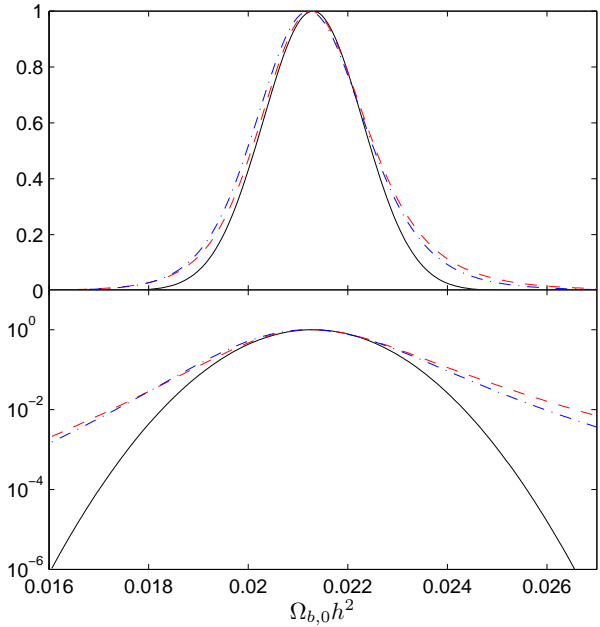}}
  \caption{
    Likelihood distributions
    for $\Omega_{\rm b,0} h^2 {\rm (BBN)}$ for different statistical models
    of the errors in the seven D/H measurements considered here
    (see Table~\ref{tab:DtoH}). 
    The solid line is a Gaussian with  
    $\Omega_{\rm b,0} h^2 = 0.0213 \pm 0.0010$ (eq.~\ref{eq:omega_b(bbn2)}).
    The other curves show the result of marginalizing over an increased 
    error variance adding to (red dashed) or multiplying (blue dot-dashed) the 
    errors on the individual D/H determinations reported in the original
    works. The additional error variance has
    a Jeffrey-like (regularized inverse amplitude) prior. 
    All errors are assumed to be uncorrelated, and the results 
    are normalized to peak at 1.
   }
   \label{fig:bayes_dist}
\end{figure}

As discussed in Section~\ref{sec:D_abund_prim}, 
a Bayesian model that increases the error bars by some 
constant and marginalizes over the possible values 
of the constant gives an error estimate on 
$\langle ({\rm D/H})_{\rm p} \rangle$ which is comparable 
to that obtained with the bootstrap method. 
However, the tails of the distributions
have quite different shapes. 
Fig.~\ref{fig:bayes_dist} shows the result of numerically 
evaluating the  distribution of $\Omega_{\rm b,0} h^2$, 
assuming Gaussian errors on the $\log {\rm (D/H)}$ measurements, 
marginalizing over the additional error variance with a 
Jeffrey-like (i.e. inverse) prior on its amplitude, 
and including the 3\% error (also assumed to be Gaussian) 
in the relation between $\eta_{10}$ and (D/H)$_{\rm p}$
(eq.~\ref{eq:dtoh_vs_eta10}).
The broad tails arise from marginalization over 
relatively large values of the added error variance, 
and fall off as a power law, rather than exponentially 
as in the bootstrap model. The result is not very sensitive 
to the inclusion or exclusion of the two most outlying data
points in Fig.~\ref{fig:DtoH}.

Using the marginalized results for 
additive additional error variance, 
the parameter constraints change somewhat
from the values given in 
sections~\ref{sec:LCDM} and \ref{sec:BBN_and_CMB_tensors}.
For example, combining the numerical baryon 
density likelihood with \textit{WMAP~5} data 
for tensor models, 
the 95\% limit on $n_{\rm s}$ 
changes from $n_{\rm s} < 1.00$ to $n_{\rm s} < 1.01$ 
due to the tail of higher $\Omega_{\rm b,0} h^2$ values that are now allowed. 
The combined tightest 
constraints change to 
$\Omega_{\rm b,0} h^2 =  0.0226\pm 0.0006$, 
$n_{\rm s} = 0.961 \pm 0.014$, with $n_{\rm s} \geq 1$ 
still just excluded at 99\% confidence and 
$r < 0.17$ at $95\%$ confidence.

In both the Bayesian and bootstrap model, 
our results depend 
critically on the assumption that the quoted
errors on the individual determinations of 
(D/H) in Table~\ref{tab:DtoH} (original and inflated) 
are statistically independent and Gaussian. 
The changes in the values of (or limits on) 
$\Omega_{\rm b,0} h^2$, 
$n_{\rm s}$, and $r$ which we have just noted when 
changing the statistical model of the errors
give an indication of the systematic error in the result 
due to the likelihood model, even when these assumptions are satisfied. 
In future, with a larger sample of measurements of (D/H) in
QSO absorbers, it may be possible to test the assumptions
that the errors are independent and Gaussian,
and perhaps identify the origin of the excess scatter in
the existing measurements. 
Such improvements would allow the primordial abundance
of deuterium to fulfil its potential in constraining cosmological
parameters.
On a different note, the Planck satellite should independently measure 
the baryon density from the CMB alone with $\sim 0.6\%$ accuracy
in a few years, 
with measurements of several acoustic peaks 
breaking the degeneracy with $n_{\rm s}$.

\section{SUMMARY AND CONCLUSIONS}

The lowest metallicity damped \lya\ systems are good
candidates for the measurement of the primordial abundance
of deuterium, not only because the gas has suffered little
astration, but also because they preferentially arise in
gas clouds with low internal velocity dispersions, facilitating
the resolution of the isotope shift of 82\,km~s$^{-1}$.
Furthermore, the high column densities involved
give detectable D\,{\sc i} absorption in many lines of the
Lyman series.
The $z_{\rm abs} = 2.61843$ DLA in the spectrum of the
bright QSO Q0913+072 is a good case in point -- we have
reported here high S/N ratio detections of six
D\,{\sc i} absorption lines from which we deduce
$\log {\rm (D/H)} = -4.56 \pm 0.04$.
The main contribution to the error in $\log {\rm (D/H)}$
is from the uncertainty in the column density of H\,{\sc i},
rather than D\,{\sc i}; it may be possible to reduce this
further in future with more sophisticated modelling of the
Lyman forest absorption that is superimposed on the
wide, damped profile of the \lya\ line.

The value of D/H we deduce for this DLA is in good
agreement with those of four out of the six QSO absorbers
previously considered to constitute the most reliable set
of such determinations.
The other two differ from the mean of the whole sample
by $\sim 2 \sigma$, and we have identified possible
reasons why their errors may have been underestimated
in the original reports. We propose that the determination
of the primordial abundance of deuterium is converging
towards the value
$\langle \log {\rm (D/H)_p} \rangle = -4.55 \pm 0.03$,
which can be considered reliable within the
68\% confidence limits.
The corresponding
$\Omega_{\rm b,0}h^2{\rm  (BBN) } = 0.0213 \pm 0.0010$
agrees within the errors with
$\Omega_{\rm b,0}h^2{\rm  (CMB) } = 0.02273 \pm 0.00062$.
Including the former as a prior in the analysis
of the \textit{WMAP~5} data from which the latter is deduced
results in a lower mean
value of the power-law index
of primordial fluctuations, from $n_{\rm s} = 0.963 \pm 0.015$ to
$n_{\rm s} = 0.956 \pm 0.013$.
The effects on other cosmological parameters deduced from
the analysis of the CMB are more modest.
Considering together the constraints available from
\textit{WMAP~5}, (D/H)$_{\rm p}$, baryon oscillations
in the galaxy distribution, and distances to Type~Ia supernovae,
we arrive at the current best estimates
$\Omega_{\rm b,0}h^2 = 0.0224 \pm  0.0005$,
$n_{\rm s} = 0.959 \pm 0.013$ ($1 \sigma$ errors),
and $r < 0.16$ ($2 \sigma$ limit) for the ratio of
the tensor and scalar power at
$k=0.002\,{\rm Mpc}^{-1}$.

Despite the long integration time, the data presented here
have not led to a significant change in
the estimate of (D/H)$_{\rm p}$ compared with the last
paper to report new observations (O'Meara et al. 2006).
This is a good sign; together with the agreement between
$\Omega_{\rm b,0}h^2{\rm  (BBN) }$ and
$\Omega_{\rm b,0}h^2{\rm  (CMB) }$, it suggests that
we have reached a stage where the primordial abundance of 
deuterium and the cosmological density of baryons are 
sufficiently well determined quantities.
There is now less urgency to increase further the number
of D/H measurements in QSO absorbers, although
a `prime' sample of only seven data points clearly
does not allow for complacency. In particular,
an expanded sample of D/H measures at high redshifts
would improve the statistical model on which
the errors on the derived cosmological parameters 
are based.
The most metal-poor DLAs remain a fertile ground
for studying early episodes of \emph{stellar} nucleosynthesis 
(e.g. Pettini et al. 2008),
and it is certainly worthwhile continuing to be aware
of their potential for further refinements of
(D/H)${\rm _p}$.

\section*{Acknowledgments}
The worked presented in this paper is based on
UVES observations made with the European Southern
Observatory VLT/Kueyen telescope at Paranal, Chile, obtained
in programme 078.A-0185(A) and from the public data archive.
We are grateful to the ESO
time assignment committee,
and to the staff astronomers at VLT
for their help in conducting the observations. 
Bob Carswell, Jim Lewis,
and Sam Rix kindly helped with various aspects of the data analysis,
and John O'Meara and Gary Steigman offered valuable 
comments on an earlier version of the paper.
We also thank the anonymous referee for useful suggestions
on the presentation of the data.
We acknowledge the use of the Legacy Archive for Microwave Background Data
Analysis (LAMBDA). Support for LAMBDA is provided by the NASA Office of
Space Science. MTM thanks the Australian Research Council for a QEII research
fellowship (DP0877998). CCS's research is partly supported by grant
AST-0606912 from the US National Science Foundation.

\newpage
\begin{appendix}
\section{UVES SPECTRUM OF Q0913+072}

For the interested reader, we provide the final 
reduced and co-added UVES spectrum of
Q0913+072, as used in the work reported here and 
in Pettini et al. (2008). Relevant details are given
in Section~2. The spectrum is available
 in its entirety in the electronic edition of 
the journal. A portion is shown here for
guidance regarding its form and content.\\

  \centering
    \begin{tabular}{@{}rrr}
      \hline
    \multicolumn{1}{c}{Wavelength$^{\rm a}$}
& \multicolumn{1}{c}{Relative Flux}
& \multicolumn{1}{c}{$ 1 \sigma$ Error}\\
\multicolumn{1}{c}{(\AA)}
& \multicolumn{1}{c}{(Arbitrary Units)}
& \multicolumn{1}{c}{in Relative Flux}\\
      \hline 
3.2999990E+03  & $-$4.8096318E+00  & 1.9800955E+01\\
3.3000266E+03  & 7.2700119E+01  & 2.1541672E+01\\
3.3000540E+03  & 8.8452042E+01  & 1.9596445E+01 \\
3.3000815E+03  & 6.0643616E+01  & 1.8316561E+01\\
3.3001091E+03  & 5.9261909E+01  & 2.0739399E+01\\
3.3001367E+03  &$-$8.5954037E+00  & 2.0203918E+01\\
3.3001641E+03  & 6.8833572E$-$01  & 1.9290850E+01\\
3.3001917E+03  & 9.1782742E+00  & 1.8983492E+01\\
3.3002192E+03  &$-$8.3213625E+00  & 2.0355886E+01\\
3.3002468E+03  & 2.1003305E+01  & 2.0274757E+01\\
      \hline
    \end{tabular}

\flushleft{\hspace{0.53cm}$^{\rm a}${Vacuum heliocentric.}
}

\end{appendix}


\begin{thebibliography}{}
\small
\itemindent -0.48cm

\bibitem[]{} Adams, T.~F.\ 1976, \aap, 50, 461


\bibitem[Akerman et al.(2005)]{} Akerman, C.~J., Ellison, S.~L.,
Pettini, M., \& Steidel, C.~C.\ 2005, \aap,  440, 499

\bibitem[Asplund et al.(2005)]{2005ASPC..336...25A} Asplund, M., Grevesse,
N., \& Sauval, A.~J.\ 2005, in
Barnes T.~ G., III, Bash F.~N., eds,
ASP Conf. Ser. Vol. 336,
Cosmic Abundances as Records of Stellar
Evolution and Nucleosynthesis,
Astron. Soc. Pac., San Francisco,  p.~25

\bibitem[Astier et  al.(2006)]{2006A&A...447...31A}
Astier, P., et al.\ 2006, \aap, 447, 31


\bibitem[Blake et al.(2007)]{2007MNRAS.374.1527B} Blake, C., Collister, A.,
Bridle, S., \& Lahav, O.\ 2007, \mnras, 374, 1527

\bibitem[Bridle et al.(2003)]{2003Sci...299.1532B} Bridle, S.~L., Lahav,
O., Ostriker, J.~P., \& Steinhardt, P.~J.\ 2003, Science, 299, 1532

\bibitem[]{} Burles, S., Tytler, D.\ 1998a, \apj, 499, 699

\bibitem[]{} Burles, S., Tytler, D.\ 1998b, \apj, 507, 732

\bibitem[Crighton et al.(2004)]{2004MNRAS.355.1042C} Crighton, N.~H.~M.,
Webb, J.~K., Ortiz-Gil, A.,
\& Fern{\'a}ndez-Soto, A.\ 2004, \mnras, 355, 1042

\bibitem[Dekker et al.(2000)]{2000SPIE.4008..534D} Dekker, H., D'Odorico,
S., Kaufer, A., Delabre, B., \& Kotzlowski, H.\ 2000,
in Masanori, I., Moorwood, A.~F.~M., eds.,
Proc SPIE, Vol. 4008,
Optical and IR Telescope Instrumentation and Detectors.
SPIE, Bellingham, p.~534

\bibitem[]{} Dunkley, J. et al.\ 2008, \apjs,  submitted (arXiv:0803.0586)

\bibitem[]{} Efron, B., Tibshirani, R.~J.\ 1993, An Introduction to the
Bootstrap. Chapman \& Hall, New York

\bibitem[Erni et al.(2006)]{2006A&A...451...19E} Erni, P., Richter, P.,
Ledoux, C., \& Petitjean, P.\ 2006, \aap, 451, 19


\bibitem[Hinshaw et al.(2008)]{2008arXiv0803.0732H} Hinshaw, G., et al.\
2008,  \apjs,  submitted (arXiv:0803.0732)

\bibitem[Hobbs(1974)]{1974ApJ...191..381H} 
Hobbs, L.~M.\ 1974, \apj, 191, 381 

\bibitem[]{} Kirkman, D., Tytler, D, Suzuki, N., O'Meara, J.~M.,
Lubin, D.\ 2003,  \apjs, 149, 1

\bibitem[]{} Komatsu, E., et al.\
2008,  \apjs,  submitted (arXiv:0803.0547).

\bibitem[Ledoux et al.(1998)]{1998A&A...337...51L} Ledoux, C., Petitjean,
P., Bergeron, J., Wampler, E.~J.,  Srianand, R.\ 1998, \aap, 337, 51

\bibitem[Ledoux et al.(2006a)]{2006A&A...457...71L} Ledoux, C., Petitjean,
P., Fynbo, J.~P.~U., M{\o}ller, P., \& Srianand, R.\ 2006, \aap, 457, 71

\bibitem[Levshakov et al.(2002)]{2002ApJ...565..696L} Levshakov, S.~A.,
Dessauges-Zavadsky, M., D'Odorico, S., \& Molaro, P.\ 2002, \apj, 565, 696

\bibitem[Lewis \& Bridle(2002)]{2002PhRvD..66j3511L}
Lewis, A., \& Bridle, S.\ 2002, Phys. Rev. D, 66, 103511

\bibitem[Linsky et al.(2006)]{2006ApJ...647.1106L} Linsky, J.~L., et al.\
2006, \apj, 647, 1106

\bibitem[]{} Molaro, P.\ 2008, in
Knapen, J.~H., Mahoney, T.~J., Vazdekis, A. eds,
ASP Conf. Ser. Vol. 390,
Pathways Through an Eclectic Universe.
Astron. Soc. Pac., San Francisco, p.~472

\bibitem[{{Morton}(2003)}]{morton03}
{Morton}, D.~C. 2003, \apjs, 149, 205

\bibitem[Murphy et al.(2007)]{2007MNRAS.376..673M} Murphy, M.~T., Curran,
S.~J., Webb, J.~K., M{\'e}nager, H., \& Zych, B.~J.\ 2007, \mnras, 376, 673

\bibitem[]{} O'Meara, J.~M., Burles, S., Prochaska, J.X., Prochter, G.~E.\ 2006,
\apj, 649, L61

\bibitem[]{} O'Meara, J.~M., Tytler, D., Kirkman, D., Suzuki, N.,
Prochaska, J.X., Lubin, D., Wolfe, A.~M.\ 2001,
\apj, 552, 718

\bibitem[Percival et al.(2007)]{2007MNRAS.381.1053P} Percival, W.~J., Cole,
S., Eisenstein, D.~J., Nichol, R.~C., Peacock, J.~A., Pope, A.~C.,
\& Szalay, A.~S.\ 2007, \mnras, 381, 1053

\bibitem[Pettini(2006)]{2006ASPC..348...19P} Pettini, M.\ 2006,
in Sonneborn, G., Moos, W., Andersson, B.-G., eds, ASP Conf. Ser. Vol. 348,
Astrophysics in the Far Ultraviolet: Five Years of Discovery with FUSE.
Astron. Soc. Pac., San Francisco, p.~19

\bibitem[]{}Pettini, M., Bowen, D.~V.\ 2001, \apj, 560, 41

\bibitem[Pettini et al.(1997)]{1997ApJ...486..665P} Pettini, M., Smith,
L.~J., King, D.~L.,  Hunstead, R.~W.\ 1997, \apj, 486, 665

\bibitem[Pettini et al.(2008)]{2008MNRAS.385.2011P} Pettini, M., Zych,
B.~J., Steidel, C.~C., Chaffee, F.~H.\ 2008, \mnras, 385, 2011

\bibitem[]{} Pontzen, A., Governato, F., Pettini, M., Booth, C.~M.,
Stinson, G., Wadsley, J., Brooks, A., Quinn, T., Haehnelt, M.\ 2008,
\mnras, in press (arXiv:0804.4474)

\bibitem[Prantzos \& Ishimaru(2001)]{2001A&A...376..751P}
Prantzos, N., \& Ishimaru, Y.\ 2001, \aap, 376, 751

\bibitem[]{} Prochaska, J.~X., Chen, H.-W., Wolfe, A.~M.,
Dessauges-Zavadsky, M., Bloom, J.~S.\ 2008, \apj, 672, 59

\bibitem[Riess et al.(2004)]{2004ApJ...607..665R} Riess, A.~G., et al.\
2004, \apj, 607, 665

\bibitem[Romano et al.(2006)]{2006MNRAS.369..295R} Romano, D., Tosi, M.,
Chiappini, C., \& Matteucci, F.\ 2006, \mnras, 369, 295

\bibitem[]{} Savage, B.~D., \& Sembach, K.~R.\ 1991, \apj, 379, 245

\bibitem[Simha \& Steigman(2008)]{2008JCAP...06..016S} 
Simha, V., \& Steigman, G.\ 2008, Journal of Cosmology and 
Astro-Particle Physics, 6, 16 

\bibitem[Steigman(2006)]{2006JCAP...10..016S} Steigman, G.\ 2006, Journal 
of Cosmology and Astro-Particle Physics, 10, 16 

\bibitem[]{} Steigman, G.\ 2007, ARNPS, 57, 463

\bibitem[Steigman et al.(2007)]{2007MNRAS.378..576S} Steigman, G., Romano,
D., Tosi, M.\ 2007, \mnras, 378, 576

\bibitem[Str{\"o}mgren(1948)]{1948ApJ...108..242S} Str{\"o}mgren, B.\ 1948,
\apj, 108, 242

\bibitem[]{} Tytler, D., Fan, X.-M., Burles, S., Cottrell, L., Davis, C.,
Kirkman, D., Zuo, L.\ 1995, in Meylan, G., ed,
QSO Absorption Lines. Springer-Verlag, Berlin, p.~289

\bibitem[Wolfe et al.(2005)]{2005ARA&A..43..861W} Wolfe, A.~M., Gawiser,
E., \& Prochaska, J.~X.\ 2005, \araa, 43, 861

\bibitem[Wood-Vasey et al.(2007)]{2007ApJ...666..694W} Wood-Vasey, W.~M.,
et al.\ 2007, \apj, 666, 694

\end{thebibliography}
\end{document}